\documentclass[letter,twocolumn]{jpsj3}
\usepackage{txfonts}
\usepackage[dvipdfmx]{graphicx}
\usepackage{color}

\title{Metamagnetic Transition in Heavy Fermion Superconductor UTe$_2$}

\author{
Atsushi Miyake$^1$\thanks{miyake@issp.u-tokyo.ac.jp}, 
Yusei Shimizu$^2$, 
Yoshiki J. Sato$^2$, 
DeXin Li$^2$, 
Ai Nakmura$^2$, 
Yoshiya Homma$^2$, 
Fuminori Honda$^2$, 
Jacques Flouquet$^3$, 
Masashi Tokunaga$^1$, and 
Dai Aoki$^{2,3}$
}

\inst{
$^1$The Institute for Solid State Physics, University of Tokyo, Kashiwa, Chiba 277-8581, Japan\\
$^2$Institute for Materials Research, Tohoku University, Oarai, Ibaraki 311-1313, Japan \\
$^3$University Grenoble Alpes, CEA, IRIG-PHELIQS, F-3800 Grenoble, France
} %\\

\abst{
We have studied the magnetization of the recently discovered heavy fermion superconductor UTe$_2$ up to 56~T in pulsed-magnetic fields.
A first-order metamagnetic transition has been clearly observed at $H_{\rm m}$~=34.9~T when the magnetic field $H$ is applied along the orthorhombic hard-magnetization $b$-axis.
The transition has a critical end point at $\sim$11~K and 34.8~T,
where the first order transition terminates and changes into a crossover regime. 
Using the thermodynamic Maxwell relation, we have evaluated the field dependence of the Sommerfeld coefficient of the specific heat directly related to the superconducting pairing.
From the analysis, we found a significant enhancement of the effective mass centered at $H_{\rm m}$, 
which is reminiscent of the field-reentrant superconductivity of the ferromagnet URhGe in transverse fields.
We discuss the origin of their field-robust superconductivity.}

\begin{document}
\maketitle

%%%**Introduction**
Heavy fermion superconductivity (SC) is one of the most interesting topics in strongly correlated electron systems (SCES). 
In particular, %SC in uranium compounds is still quite fascinating in spite of a long history after the first discovery in UBe$_{13}$ in 1983~\cite{Ott83}.
the discovery of coexistence of ferromagnetism (FM) and spin triplet SC in UGe$_2$ opens a new chapter~\cite{Sax00}.
%Many interesting phenomena are already known, such as the coexistence of antiferromagnetism and SC in UPd$_2$Al$_3$~\cite{Gei91}, the multiple SC phases in UPt$_3$~\cite{Joy02} and (U,Th)Be$_{13}$~\cite{Hef90}, the chiral $d$-wave SC in URu$_2$Si$_2$ coexisting with the hidden order~\cite{Kit16}.
%Furthermore, the discovery of coexistence of ferromagnetism (FM) and spin triplet SC in UGe$_2$ opens a new chapter~\cite{Sax00}.
The appearance of new materials such as URhGe~\cite{Aok01} and UCoGe~\cite{Huy07} with low Curie temperature ($T_{\rm Curie}$) shows the key role of Ising interaction and was hints on elegant way to obtain field-reentrant SC or its reinforcement by lowering $T_{\rm Curie}$ in transverse field scan with respect to the initial perpendicular FM sublattice magnetization~\cite{Levy2005,Aoki2009,Aok12_JPSJ_review,Aoki2019re}.
%A proposed new perspective is that these systems may be a good platform for topological SC as well.

\begin{figure}
\begin{center}
\includegraphics[width=\hsize]{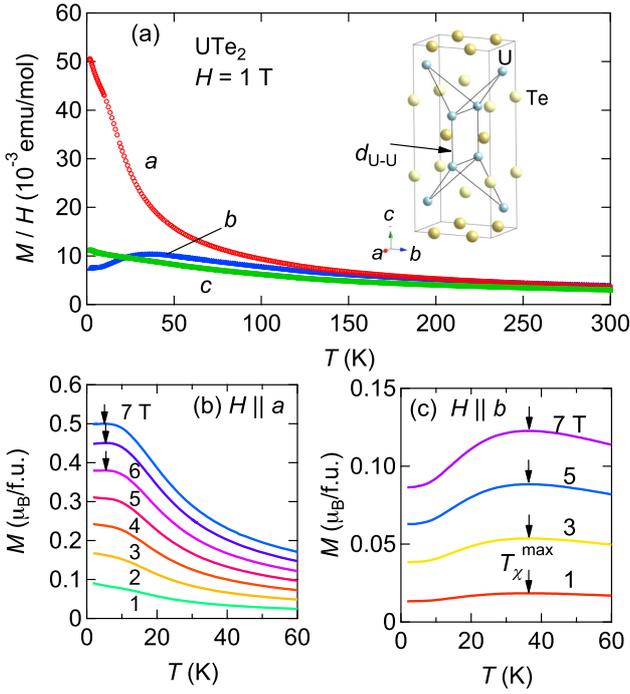}
\end{center}
\caption{(Color online) (a) Temperature dependence of $M/H$ at 1~T for $H~||~a$, $b$, $c$-axes. 
The inset in (a) shows the crystal structure of UTe$_2$.
$M(T)$ curves at different fields for $H~||~a$ and $b$ are shown in (b) and (c), respectively.
%\red{For $H~||~a$, the fields varies between 1 and 7~T by 1~T step.}
The arrows indicate the maximum of $M(T)$.
}
\label{MT}
%\vspace{-0.5cm}
\end{figure}

Recently, SC was discovered in the paramagnetic (PM) UTe$_2$, with a relatively high superconducting transition temperature, $T_{\rm sc}$~=~1.6~K~\cite{Ran2018,Aoki2019}.
SC is believed to be of unconventional spin-triplet type, since the $H_{c2}$ highly exceeds the Pauli limit.
In contrast to the previous spin-triplet FM superconductors, the ground state of UTe$_2$ is PM at least down to $T_{\rm sc}$, most probably locating near a FM instability~\cite{Ran2018}.
UTe$_2$ crystallizes in a body-centered orthorhombic structure (space group: $Immm$) [see the inset of Fig~\ref{MT}(a)].
The nearest-neighbor (NN) U atoms in UTe$_2$ align along the $c$-axis with the shortest U-U distance $d_{\rm U-U}=3.78$~\AA~\cite{Ikeda2006}.
A similarity with the previous FMSC is the field robust SC phase for $H~||~b$-axis, which is perpendicular to the easy magnetization $a$-axis at low fields.
The SC phase still exists even at 20~T~\cite{Ran2018}, although the temperature dependence of $H_{\rm c2}$ appears sample dependent~\cite{Aoki2019}.
It is expected that magnetic fluctuations will drastically develop at high fields, being coupled with the Fermi surface instabilities when the magnetic polarization reaches a sufficiently high critical value.

A key observation in UTe$_2$ for $H~||~b$-axis is a maximum of the magnetic susceptibility $\chi$ at $T_{\chi}^{\rm max}\sim 35$~K~\cite{Ran2018,Ikeda2006}.
In metallic SCES, many compounds show metamagnetic behavior at $H_{\rm m}$, which have a similar energy scale to their $T_{\chi}^{\rm max}$~\cite{Aoki2013, Knafo2012}.  
It is also interesting to mention the expected magnetic anisotropy from the point of view of the crystal structure.
In many U compounds, the magnetic moments are aligned along a direction perpendicular to the axis connecting the NN U atoms in the magnetic ordered state~\cite{Robinson1994}.
The NN U atoms of URhGe and UCoGe, for example, make a zigzag-chain along the $a$-axis, resulting in a FM ground state with the easiest magnetization $c$-axis~\cite{Aoki2019re}.  
By applying transverse magnetic fields ($H~||~b$-axis), reentrant SC in URhGe~\cite{Levy2005} and field-reinforced SC  in UCoGe~\cite{Aoki2009} occur.
The enhancement of the FM fluctuations in the $bc$-plane is considered to have an important role~\cite{Aoki2019re}.
For UTe$_2$, the development of magnetic fluctuations in the $ac$-plane is expected at high fields. 

Here, we performed magnetization ($M$) measurements in UTe$_2$ for $H~||~a$ and $b$-axes in pulsed-magnetic fields up to 56~T and found a first order metamagnetic transition (MMT) accompanied by a huge jump of $M$ ($\Delta M\sim 0.6\,\mu_{\rm B}/{\rm f.u.}$) for $H~||~b$-axis at $H_{\rm m}\sim 34.9\,{\rm T}$.
%The transition is clearly of first order., although the hysteresis loop is very small.  
We also observed a small anomaly in the $M(H)$ curve for $H~||~a$-axis near 6.5~T.
The MMT seems to occur when $M$ for $H~||~b$ reaches that for $\mu_0H\sim6.5$~T along the $a$-axis.  
We discuss the $H$-dependence of the Sommerfeld coefficient ($\gamma$) derived from the temperature dependence of $M$ using a thermodynamic Maxwell relation, and 
the $H$-robustness of $T_{\rm sc}$ triggered by the enhancement of the effective mass.
 
%%%**Experiment**

Single crystals of UTe$_2$ were grown using the chemical vapor transport method~\cite{Ran2018, Aoki2019}.
%The starting materials of U and Te were put into the quartz ampoule with the atomic ratio U~:~Te~=~2~:~3 together with iodine as the transport agent. 
%The ampoule was heated slowly up to 900~$\degC$ for pre-reaction, and then the temperature gradient was applied (1050/1000~$\degC$) for a few weeks in a horizontal furnace. 
%Many single crystals with a few millimeter size were obtained. 
%The single crystal was checked by Laue photographs and single crystal X-ray analysis.
$M$ in pulsed-magnetic fields was measured by the conventional induction method, employing coaxial pick-up coils.
Pulsed-magnetic fields up to 56~T were applied using nondestructive pulse-magnets having typical durations of $\sim 36$~ms installed at the International MegaGauss Science Laboratory of the Institute for Solid State Physics of the University of Tokyo.
%\red{We confirmed SC transition through the magnetization measurements, indicative of the high quality of sample used here \cite{suppl}.}
The measurements were done for the field applied along the $a$ and $b$-axes and at low temperatures down to 1.4~K.
Below 7~T, the temperature dependence of $M$ was measured by a commercial SQUID magnetometer at temperatures down to 1.8~K for $H~||~a, b, c$-axes.

%%%*******results and discussion*****
%---------------------
%---------------------
Figure~\ref{MT}(a) shows temperature ($T$) dependence of the magnetic susceptibility $M/H$ with $H$ applied along the orthorhombic principal axes. 
Consistently with previous reports~\cite{Ran2018, Ikeda2006}, there is no indication of any phase transitions down to 1.8~K, suggesting a PM ground state before the establishment of SC.
$M(T)$ for $H~||~b$-axis shows a broad maximum at $T_{\chi}^{\rm max}\sim$~36~K, which hardly depends on $H$ at least up to 7~T [Fig.~\ref{MT}(c)].
On the other hand, as shown in Fig.~\ref{MT}(b), $M(T)$ for $H\parallel a$ increases upon cooling without saturation at low fields, 
but $M(T)$ tends to saturate around 5~K above $\sim$5~T.
These saturations may correspond to the anomaly around $7\,{\rm T}$ in $M(H)$ curve shown later.

%---------------------
\begin{figure}
\begin{center}
\includegraphics[width=\hsize]{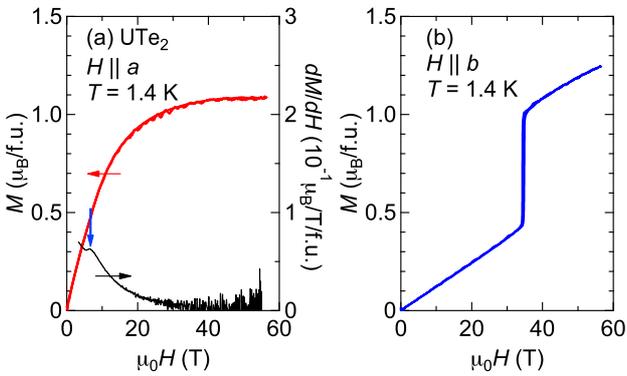}
\end{center}
\caption{(Color online) Magnetization curves of UTe$_2$ at 1.4~K for (a) $H~||~a$-axis and (b) $H~||~b$-axis.
The differential susceptibility $dM$/$dH$ is also shown for $H~||~a$-axis.}
\label{MH}
%\vspace{-0.5cm}
\end{figure}
\begin{figure}
\begin{center}
\includegraphics[width=\hsize]{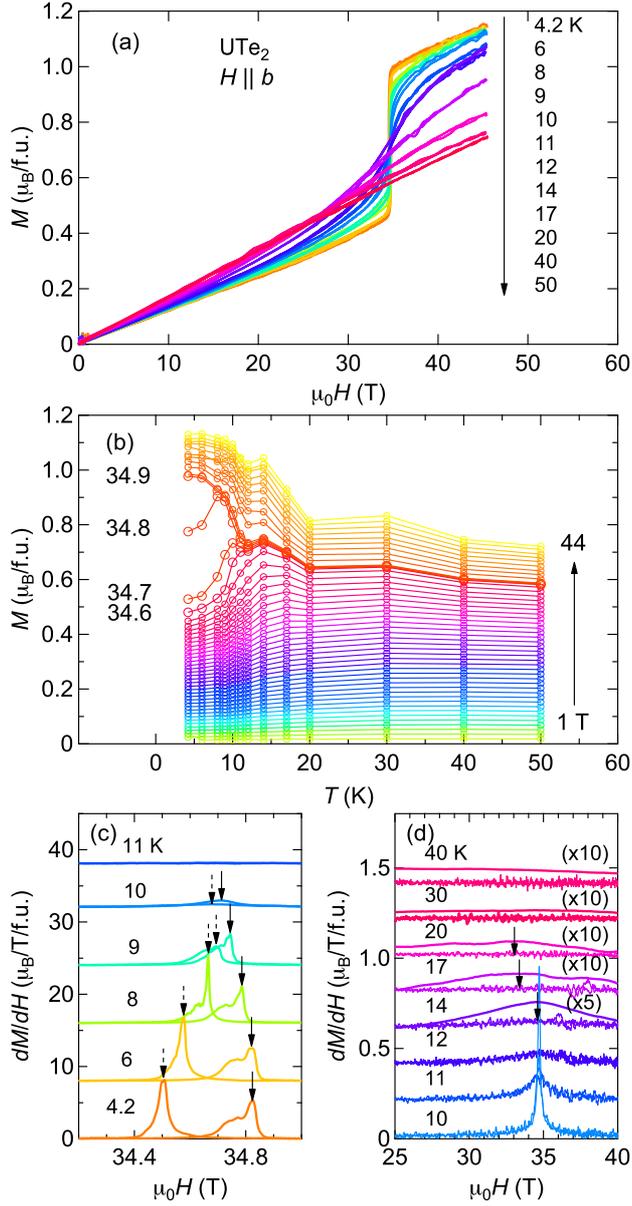}
\end{center}
\caption{(Color online) (a) $M(H)$ curves at various temperatures for $H~||~b$-axis in UTe$_2$. 
(b)Temperature dependence of magnetization at various constant fields from 1 to 44~T with 1~T step for the field up-sweep data shown in (a).
Near the CEP between 34.6 and 34.9~T, $M(T, H)$ curves are shown with the step of 0.1~T. 
Field dependence of $dM$/$dH$ in UTe$_2$ for $H~||~b$-axis at (c)low and (d)high temperatures. 
For clarity, the data are offset by (c) 8 and (d) 0.2 $\mu _{\rm B}$/T/f.u.
The solid and broken arrows in (c) indicate the $H_{\rm m}$ for up-sweep and down-sweep measurements.
The solid lines superimposed on several data in (d) are the magnified $dM$/$dH$ curves, which are smoothed strongly for clarity. 
}
%\vspace{-0.5cm}
\label{MH_MT}
\end{figure}

%---------------------
%Most stimulating observation in this study is a field-induced MMT at $H_{\rm m}$~=~34.8~T with applied fields along the $b$-axis.
Next, we show the results of $M(H)$ in a pulsed field.
The results at low field and obtained $H_{c2}(T)$ curves are shown in the Supplemental Material~\cite{suppl}.
Figure~\ref{MH} presents $M(H)$ curves of UTe$_2$ for $H~||~a$- and $b$-axes at 1.4~K.
For the easiest magnetization $a$-axis at low fields, $M$ monotonically increases and tends to saturate with increasing fields.
Note that a small anomaly around 7~T is observed, as seen in the differential susceptibility $dM$/$dH$.
Remarkably, a huge jump in $M$ ($\Delta M\sim 0.6~\mu_{\rm B}$/f.u.) appears for $H~||~b$-axis at $H_{\rm m}=34.9$~T, accompanied by a clear but rather small hysteresis loop.
It is noted that $\Delta M$ slightly decreases after several thermal and field cycles without any changes of $H_m$ and the slope of $M(H)$.
It may arise from the strong field angle dependence or the damage due to the magnetostriction.
The data shown in Fig.~\ref{MH_MT} between 4.2~K and 17~K, which are used for the derivation of Sommerfeld coefficient in the specific heat $\gamma$, are obtained in a same run.
The hysteresis is clearly seen in $dM/dH$ in Fig.~\ref{MH_MT}(c).
These results indicate that the MMT in UTe$_2$ is of first order.
A nonzero intercept of a linear extrapolation of $M(T, H$$\to$0) from $H>H_{\rm m}$ suggests that a finite ordered moment exists in the field-induced phase;
as if there is a switch from the PM to FM phase at $H_{\rm m}$.
Furthermore, the remaining large slope of $M$ $dM/dH\sim 0.01~\mu_{\rm B}/{\rm T}$ above $H_{\rm m}$ is a mark of surviving heavy quasiparticles in the polarized FM phase.
Clearly, the $b$-axis becomes the easy-magnetization axis above $H_{\rm m}$.
As for URhGe, the magneto-crystalline effect changes drastically with the field.
Interestingly, the value of $M$ occurring the metamagnetic transition for $H\parallel b$ is close to that
showing an anomaly for $H\parallel a$-axis, namely $M\sim 0.5~\mu_{\rm B}/$f.u. 
It is quite common in heavy fermion system that for a critical value of $M$, 
Fermi surface instabilities associated with enhancement of the magnetic fluctuations occur~\cite{Hardy2011, Aoki2019re}. 

%-----------------------------
%-----------------------------
In order to know the $T$ evolution of the MMT, we measured the $M(H)$ curves at different temperatures, as shown in Fig.~\ref{MH_MT}(a).
$\Delta M$ becomes smaller with increasing $T$.
Interestingly, the value of $H_{\rm m}$ is almost $T$-independent, in stark contrast to the typical itinerant metamagnets locating near the FM critical point~\cite{Bra16}. 
Figure~\ref{MH_MT}(b) presents the $M(T)$ curves at constant various fields replotted from the results in Fig.~\ref{MH_MT}(a).
The low-field $M(T)$ with a broad maximum at $T_{\chi}^{\rm max}$ changes into a rapid increase of $M(T)$ on cooling above $H_{\rm m}$.
There exists a critical end point (CEP) around $11\,{\rm K}$ and $34.7\,{\rm T}$ with a sign change of the $M(T)$ slope.
The signature of the CEP is further confirmed by the differential susceptibility $dM$/$dH$ in Figs.~\ref{MH_MT}(c) and \ref{MH_MT}(d), which will be discussed later.
%Across $H_{\rm m}$, the huge peak of d$M$/d$H$ is observed with a relatively small $H$ hysteresis $\Delta H$, i.e., $\Delta H\sim$~0.32~T.
%$H_{\rm m}$ shows a weak temperature dependence accompanied by the reduction of $\Delta H$ with increasing temperature.
%Above $\sim$11~K, the hysteresis disappears, and the peak becomes very broad, indicating that the first order MMT changes to a crossover-like behavior.
The field-switch from PM to FM for $H\parallel b$-axis resembles to that observed in FM compounds such as UGe$_2$
with the detection of FM wings in ($T,P,H$) phase diagram.~\cite{Tau10,Kot11,Bra16}.
The significant differences here are: 
i) the PM-FM line, $H_{\rm m}(T)$ is weakly temperature-dependent in UTe$_2$, 
ii) the phenomena occur for a transverse field with respect to the initial easy-magnetization axis in UTe$_2$,
contrary to the case in UGe$_2$ where the FM wing structure is observed for the longitudinal field scan~\cite{Tau10,Kot11}.
%Further consequence of the critical point is a sudden change of temperature dependence of magnetization at critical point in Fig.~\ref{MH}(b).

%At higher temperatures, the MMT becomes crossover.
%It is also noted that SC to normal state is detected through the magnetization measurements, indicative of the quality of sample used this study.   

%------------------
\begin{figure}
\begin{center}
\includegraphics[width=\hsize]{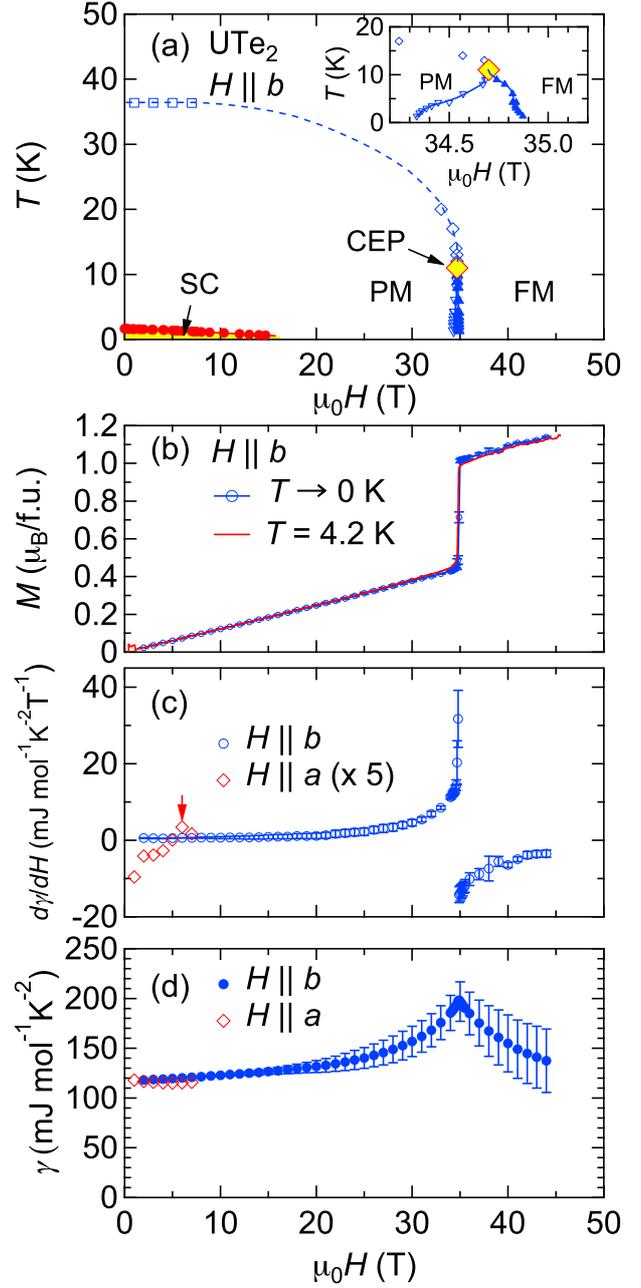}
\caption{(Color online)(a) Magnetic phase diagram of UTe$_2$ for $H~||~b$-axis. 
The inset focuses near the critical end point CEP.
The upward (downward) triangles, diamonds, circles and squares correspond to the $H_{\rm m}$ for up-sweep (down-sweep), $H_{\rm m}$ above CEP, $H_{c2}$~\cite{Aoki2019} and $T_{\chi}^{\rm max}$, respectively.   
The dotted line is a guide to the eye.  
(b) $M(H)$ curves of UTe$_2$ for $H~||~b$-axis for $T=4.2\,{\rm K}$ (solid line) and $T\to 0\,{\rm K}$ (symbol and line) obtained from an extrapolation assuming the relation $M(T) =M_0 +\beta T^2$ based on a Fermi liquid behavior.
Field dependence of (c) $d\gamma$/$dH$ and (d) $\gamma$ for $H~||~a$ and $b$ axes derived from the thermodynamic Maxwell relation.
An arrow in (c) indicates a peak of $d\gamma$/$dH$ for $H~||~a$-axis.
%$\gamma(0~{\rm T})=118$~mJ~mol$^{-1}$K$^{-2}$ and the $\gamma_0=61$~mJ~mol$^{-1}$K$^{-2}$ [the horizontal line in (b)] were the reported values in Ref.~\citen{Aoki2019}.  
}
\label{HTPD}
\end{center}
%\vspace{-0.5cm}
\end{figure}
%------------------

From the $M$ measurements shown in Figs.~\ref{MH}(b) and \ref{MH_MT}, the magnetic phase diagram in UTe$_2$ for $H~||~b$-axis is summarized in Fig.~\ref{HTPD}(a).
At 1.4~K, the $H$-induced MMT takes place at $H_{\rm m}$~=~34.9~T for $H$-up sweep measurements.
With increasing temperature, $\Delta H$ decreases and is suppressed to zero at the CEP ($11\,{\rm K}$ and $34.7\,{\rm T}$).
Above the CEP, the crossover-like broad maximum in $dM$/$dH$ appears at $H_{\rm m}$,
and $H_{\rm m}$ decreases with increasing temperature.  
Since the anomaly in $dM$/$dH$ becomes indiscernible above 30~K.
%, it is not clear how $H_{\rm m}$ varies at higher temperatures.
Note that the extrapolation to the higher temperature seems to be connected to $T_{\chi}^{\rm max}$ at low fields.
These results agree with the scaling of $H_{\rm m}$ and $T_{\chi}^{\rm max}$~\cite{Aoki2013}. 
Thus the MMT and $T_{\chi}^{\rm max}$ are dominated by a same single energy scale. 
We also note that the phase diagram is consistent with the results obtained from magnetoresistance~\cite{Knafo}.

In order to clarify the evolution of the electronic states through the MMT, we have analyzed the $M(T)$ data [Fig.~\ref{MT} (b) for $H~||~a$ and Fig.~\ref{MH_MT} (b) for $H~||~b$] using a thermodynamic Maxwell relation following the previously successful reports, such as in heavy fermion metamagnet CeRu$_2$Si$_2$~\cite{Paulsen1990, Sakakibara1995} and the reentrant FMSC URhGe~\cite{Hardy2011}.
The relation between $M$ and the entropy $S$ is known from the Maxwell relation: 
$\left(\partial S/\partial H\large\right)_T = \left(\partial M/\partial T\large\right)_H$.
At low temperature, $S = \gamma T$ was observed in UTe$_2$~\cite{Ran2018,Aoki2019}.
From these relations, we can directly access the $H$ dependence of $\gamma$ as,
$\left(\partial \gamma/\partial H\large\right)_T = \left(\partial^2 M/\partial T^2\right)_H$.
Using the $M(T)$ data at various temperatures and assuming that $M(T)$ varies as $T^2$ at low $T$ on the basis of the Fermi liquid state, we have evaluated $\gamma$ as a function of $H$.
Figure~\ref{HTPD} (b) shows the $M(H)$ curves at 4.2~K and 0~K.
The latter was derived by extrapolation to 0~K assuming $M(T, H)=M(0, H)+\beta T^2$, where $\beta$ is a coefficient of $T^2$-dependence of $M$.
As expected, $M$ at below (above) $H_{\rm m}$ is smaller (larger) at lower $T$, indicating the reliability of assumption of the $T^2$-dependence. 
As shown in Fig.~\ref{HTPD} (c), $d\gamma$/d$H$ for $H~||~a$ and $b$ axes is evaluated using $d\gamma$/$dH=2\beta$.
In both $H$-directions, a peak structure is observed at 6~T and 34.8~T for $H~||~a$ and $b$. 
The former is in agreement with $dM/dH$ anomaly [see Fig.~\ref{MH}(a)].
Remarkably, for the $b$-axis a very sharp singularity with a drastic sign change is observed at $H_{\rm m}$.
The change in sign of $d\gamma$/d$H$ takes place within an analyzed $H$-step of 0.1~T, reflecting a sharp $M$ jump at $H_{\rm m}$.
The $\gamma(H)$ derived by integrating $d\gamma$/d$H$ is shown in Fig.~\ref{HTPD} (d).
The reported value of $\gamma$=~118~mJ~mol$^{-1}$K$^{-2}$ is used for $\gamma(0)$~\cite{Aoki2019}.  
Of course as the MMT at $H_m$ is of first order, discontinuity in $\gamma(H)$ can exist at $H_m$.
The choice of a converging value at $H_m$ coming from low field ($H<H_m$) or high field ($H>H_m$) scan is in agreement with the singularity in the field dependence of the coefficient of $T^2$-term of the resistivity $A(H)$~\cite{Knafo}, which often scales $\gamma^2$ in heavy fermion compounds.
$\gamma(H)$ for $H\parallel a$ weakly depends on $H$ and shows a small minimum at 6~T, which are in agreement with the $A(H)$~\cite{Aoki2019}.
As seen in other heavy fermion metamagnets, a clear peak in $\gamma(H)$ centered at $H_{\rm m}$ is observed for $H~||~b$-axis.
This fact strongly suggests the development of fluctuations on approaching to $H_{\rm m}$.
The singularity of $\gamma (H)$ at the first order transition $H_{\rm m}$ 
demonstrates that the dynamic correlation survives through $H_{\rm m}$ in agreement with recent work on so called quantum annealed criticality~\cite{Cha18}
%--------------------
\begin{figure}
\begin{center}
\includegraphics[width=\hsize]{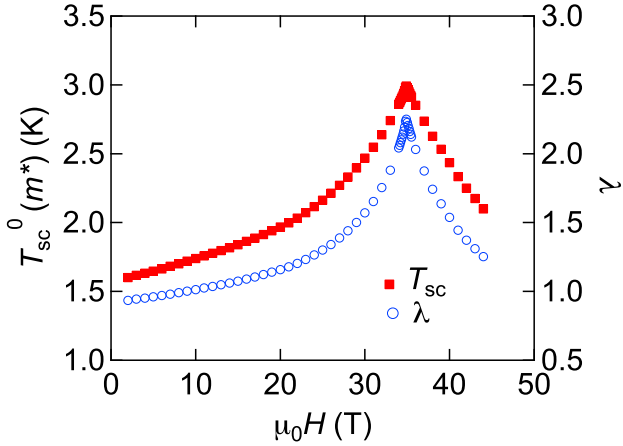}
\caption{(Color online) Field dependence of the simulated $T_{\rm sc}$ and $\lambda$ of UTe$_2$ for $H~||~b$-axis. 
}
\label{lambda}
%\vspace{-0.5cm}
\end{center}
\end{figure}
%--------------------

Finally, we discuss the possibility of the $H$-reinforced SC in UTe$_2$ on going to $H_{\rm m}$ for $H~||~b$ based on a simple mass enhancement mechanism, which is proposed for the case of the reentrant SC in URhGe~\cite{Miyake2008, Miyake2009, Hardy2011, Aoki2019re}.
In similarly to UTe$_2$ shown in Fig.~\ref{HTPD}(d), a mass enhancement was observed in URhGe through the $A(H)$~\cite{Miyake2008} and $\gamma(H)$~\cite{Hardy2011} coefficients for $H$ applied hard magnetization $b$-axis: a peak appears at $H_{\rm R}\sim$~12~T, where FM magnetic moment reorients from the $c$- to $b$-axis, and reentrant SC becomes more stable than the low-field SC.
The total effective mass $m^{\ast}(H)$ can be expressed as, $m^{\ast}(H) = m^{\ast\ast}(H)+m_{\rm B}$, where $m^{\ast\ast}(H)$ and $m_{\rm B}$ are the correlation and band masses, respectively.
As seen in Fig.~\ref{HTPD} (d),  $m^{\ast}\sim\gamma(H)$ is strongly enhanced at $H_{\rm m}$.
If the band structure is assumed to be independent of $H$, namely $m_{\rm B}\sim \gamma_0$ and $\gamma_0 = 61$mJ~mol$^{-1}$K$^{-2}$~\cite{Aoki2019}, the development of the fluctuation corresponds to the enlargement of the $m^{\ast\ast}$. 
Thus, the pairing coupling constant $\lambda \equiv \frac{m^{\ast\ast}}{m_{\rm B}}$ is indicated to become larger with $H$ along the $b$-axis~\cite{Aoki2019, Ran2018}.
Assuming $m^{\ast\ast}(H) \sim \gamma (H) -\gamma_0 $, $\lambda(H)$ is derived as shown in Fig.~\ref{lambda}. 
$\lambda(H)$ increases with $H$, in agreement with the previous results~\cite{Ran2018, Aoki2019}.
Moreover, it does not tend to saturate before $H_{\rm m}$.
We have estimated $T_{\rm sc}$, employing a simplified McMillan-type formula, $T_{\rm sc}=T_0\exp\left({-\frac{\lambda + 1}{\lambda}}\right)$, where $T_0$ is a constant and is determined as the experimental $T_{\rm sc}$ at 0~T~\cite{Miyake2008, Miyake2009, Aoki2019re}.
As shown in Fig.~\ref{lambda}, $T_{\rm sc}$ becomes nearly doubled at $H_{\rm m}$.
Although this model is too rough to describe realistic systems, we believe that it will stimulate further experimental and theoretical studies. 
%Following the recipe, we evaluated the field variations of the coupling constant $\lambda$ and $T_{\rm sc}$.

%Especially for metamagnets, it is known to show a peak in the effective
%The field evolution of the electronic specific coefficient $\gamma$ can be evaluated by using the thermodynamic Maxwell relation \cite{Hardy2011}:

%It is very interesting to investigate whether the SC phase still exist at higher fields. 
%Further spectacle event is that the SC phase is reinforced by the magnetic fluctuation accompanied by the metamagnetic transition.  

Fermi surface instabilities seem to occur in UTe$_2$ for $H\parallel a$ and $b$ crossing a critical value of the magnetization. 
%This change is coupled with strong magnetic fluctuation which obviously survives despite the first order nature of the transition at $H_{\rm m}$.
%The enhancement by a factor near 2 of the zero field value of the Sommerfeld coefficient gives the correct order of magnitude to drive SC up to $H_{\rm m}$.
A new ingredient, compared to the previous FMSC cases of UCoGe and URhGe, is that SC appears in the PM ground state.
Thus, arguments used on the field increase of the SC coupling constant $\lambda$ in transverse field-scan due to the suppression of $T_{\rm Curie}$ with field appears to be not relevant here, because no FM has been detected in UTe$_2$ so far. 
A simple image one can consider is that,
on cooling the Fermi surface is fully established (below $T_{\chi}^{\rm max}$) and strong Ising fluctuations occur along the $a$-axis in the low fields.
A field applied along $b$-axis lead to drastic change of the orientation of fluctuations as observed in URhGe.

Metamagnetism in strongly correlated FM systems can have various origins.~\cite{Ima10}
For example, as emphasized earlier,
the reentrance of FM just above a critical pressure $P_c$ in longitudinal field scan is shown in UGe$_2$~\cite{Tau10,Kot11}.
The collapse of $T_{\rm Curie}$ in transverse field scan accompanied with a switch of magnetocrystalline energy is shown in URhGe~\cite{Levy2005}.
A field-controlled valence transition~\cite{Wat08} and a Fermi surface reconstruction produced by relative Zeeman decoupling of the subbands~\cite{She18} can be invoked.

A fascinating road is to go deeper in band structure calculation. 
The first LDA calculation gives the idea that UTe$_2$ will be a Kondo semiconductor; 
a shift of the 5$f$ level can restore the experimental fact that UTe$_2$ is a rather good metal at zero field~\cite{Aoki2019}. 
The magnetic field is an elegant parameter to act on Fermi surface through the change of the 5$f$ configuration and also the valence. 
Our result must push to progresses in field and pressure effects on UTe$_2$, aiming to find the striking difference in the magnetic fluctuations at a similar magnetic polarization along the $a$ and $b$-axes when $M$ reaches $0.5\,\mu_{\rm B}$/f.u.

In summary, we have studied magnetization in pulsed-fields up to 56~T for the easy- (hard-) magnetization $a$ ($b$) axis of heavy fermion superconductor UTe$_2$.
A sharp magnetization jump due to the first-order metamagnetic transition is observed for $H~||~b$-axis at $H_{\rm m}$~=~34.9~T.
In addition, a small anomaly in the $M(H)$ curve for $H~||~a$-axis is observed  at $H\sim$~6.7~T.
From careful temperature and field scans of the magnetization, we find that the the first order character terminates  at the critical end point ($\sim$~11~K, 34.8~T).
A singularity of the metamagnetic transition is of first order, 
which is originally induced from the PM ground state. 
We also propose a possible $H$-reinforcement of superconductivity, as indicated by the field dependence of the effective mass derived by a thermodynamic Maxwell relation.
UTe$_2$ provides a playground to study the interplay of SC, metamagnetism and Fermi surface instabilities.
%The magnetostriction and magnetocaloric effects will give important knowledge to reveal the origin of the metamagnetic transition, and thus of the superconductivity. 
We finally note that the metamagnetic transition in UTe$_2$ was also observed by Knafo {\it et al}~\cite{Knafo}. and Ran {\it et al}~\cite{Ran}.

\begin{acknowledgment}
We thank K. Ishida, S. Kitagawa, Y. Yanase, Y. Tokunaga, Y. Tada, W. Knafo, G. Knebel, A. Pourret, J. P. Brison, D. Braithwaite, I. Sheikin, H. Harima, and S. Ran for fruitful discussion.
This work was supported by ERC starting grant (NewHeavyFermion), KAKENHI (JP15H05884, JP15H05882, JP15K21732, JP16H04006, JP15H05745, JP19H00646), and ICC-IMR.
\end{acknowledgment}

%\appendix
%\section{}
%
%Use the \verb|\appendix| command if you need an appendix(es). The \verb|\section| command should follow even though there is no title for the appendix (see above in the source of this file).
%
%For authors of Invited Review Papers, the \verb|profile| command is prepared for the author(s)' profile.  A simple example is shown below.
%
%\begin{verbatim}
%\profile{Taro Butsuri}{was born in Tokyo, Japan in 1965. ...}
%\end{verbatim}

\end{document}